# Metabolomics of Aging and Alzheimer's Disease: From Single-Omics to Multi-Omics

Yiming Li[1], Yuan Luo[1,*]

[1]Department of Preventive Medicine, Northwestern University, USA
[*]Corresponding author: yuan.luo@northwestern.edu

**Abstract**: Aging is a multifactorial process and a key factor of morbidity and mortality. Alzheimer's disease (AD) is an age-related disorder and a main cause of worldwide disability. Both aging and AD can be characterized by metabolic dysfunction. Metabolomics can quantify the complete set of metabolites in a studied sample and is helpful for studying metabolic alterations in aging and AD. In this review, we summarize the metabolomic changes regarding aging and AD, discuss their biological functions, and highlight their potential application as diagnostic biomarkers or therapeutic targets. Recent advances in multi-omics approaches for understanding the metabolic mechanism of aging and AD are also reviewed.

## 1. Introduction

Aging is a natural process involving a chronological decline in different physiological and metabolic functions [1, 2]. Metabolic alterations including impaired mitochondrial function and deregulated nutrient sensing are among the hallmarks of aging [1, 3-5]. Since aging is a major driver of morbidity [6, 7] and mortality [8, 9], understanding the metabolic pathways associated with its mechanism will be useful for precision medicine applications [5]. Alzheimer's disease (AD) is an age-related disorder and a major cause of worldwide disability and mortality [10]. AD is recognized as a global public health priority, and further insights into AD pathogenesis is needed for the treatment and prevention of AD [10]. Previous studies have shown widespread metabolic perturbation as a defining feature of AD [11-13]. Metabolomics is a high-throughput omics technology that allows researchers to measure the complete set of metabolites in a given sample and study the alterations in different metabolism functions [14]. It is a powerful tool for studying the mechanism of aging and enables researchers to systematically study the metabolic dysfunction in AD patients, facilitating biomarker identification, drug discovery, and clinical intervention for AD. Additionally, integrating metabolomics and other omics data with phenotypic data may help us better characterize the unique health signatures of different patients [15-18]. In this review, we begin with an introduction to major analytical platforms of metabolomics and common approaches in metabolomic data analysis. Next, we discuss applications of metabolomics in aging and AD research, highlighting both the metabolomic links and divergences between natural aging and AD. Finally, we review recent multi-omics studies related to aging and AD with a focus on the metabolomic findings, and highlight new avenues of research to further explore the metabolomics of aging and AD.

## 2. Metabolism and metabolomics

Metabolism refers to the set of life-sustaining chemical reactions within an organism's cells, and metabolites are the intermediates and products of metabolism. With technological advancements, we can now identify, quantify, and characterize small biological molecule (< 1,500 daltons) metabolites found within a specific biological sample [19], and this complete set of metabolites is called the metabolome.

Mass spectrometry (MS) and nuclear magnetic resonance (NMR) spectroscopy are two analytical platforms frequently used to measure the metabolome. MS ionizes the chemical species in a sample and visualizes a mass spectrum of the ion signals sorted according to their mass-to-charge ratio [20]. It can identify metabolites rapidly but requires prior sample separation using chromatography techniques including gas chromatography [21], liquid chromatography [22], and capillary electrophoresis [23]. NMR spectroscopy detects the NMR signals produced by spinning the sample inside a magnetic field and hence may identify metabolites with different molecular structures [24]. Compared with MS, NMR spectroscopy is non-destructive because it does not require sample separation [25]. However, it is less sensitive in detecting low molecular weight molecules [26]. For a comprehensive analysis of the metabolome, parallel application of both platforms is often needed [27].

Metabolomic data analysis seeks to identify metabolites and their underlying pathways significantly associated with certain diseases or conditions. This requires harnessing high-dimensional, complex datasets – currently, the Human Metabolome Database contains 248,047 metabolite entries [28], and a single metabolomic dataset may contain hundreds of features [29, 30]. Therefore, dimensionality reduction methods like principal component analysis (PCA) [31], hierarchical clustering [32], non-negative matrix factorizations [33-36] and tensor factorizations [37-39] are often used to identify clusters of functionally related metabolites, followed by pathway analysis that studies the interactions among genes and metabolites through enrichment or topological analysis [40]. On the other hand, the high dimensionality nature of metabolomic data can be directly leveraged by machine learning methods like random forests [41], extreme gradient boosting [42], and deep learning [43-45] to characterize metabolic signatures for different diseases. The application of metabolomics may help researchers identify disease biomarkers for clinical use and better understand the etiology and pathogenesis of different disorders [14, 27].

## 3. Metabolomics of aging and Alzheimer's Disease

An organism's lifespan is known to be influenced by mitochondrial and endocrine function, nutrient sensing, and redox homeostasis [1, 3, 4]. Metabolomics allows researchers to investigate complex metabolomic variations representing changes in different metabolism functions and hence is a powerful tool for studying the mechanism of aging. Recent applications of metabolomics in aging research have been reviewed by Adav and Wang [2]. Major metabolomic biomarkers of aging identified by previous studies include amino acids such as leucine [46, 47], glycine [48, 49], and histidine [50, 51], lipids-related metabolites like triglyceride [52, 53], sphingolipids [50, 54], phospholipids [51, 55], and lipoprotein size [52, 53], as well as redox-related metabolites such as nicotinamide adenine dinucleotide (NAD) [56] and reactive oxygen and nitrogen species (RONS) [57]. These results indicate that amino acid metabolism, lipid metabolism, and redox reactions may play an important role in aging and longevity [1, 58].

Neuronal activity requires a large amount of energy, and the brain's metabolic consumption can represent 20% of the whole-body oxygen uptake [59]. A decline in brain metabolism can contribute to cognitive impairment [60], and is one of the earliest symptoms in AD [11]. To systematically study the metabolic perturbations in AD, metabolomic approaches have been widely applied, and recent advances in this field have been reviewed by Wilkins and Trushina [27]. Previous studies identified a variety of metabolomic features associated with AD, for instance, RONS [61], metabolites of the glycolytic pathway [12, 62-64], amino acids like glutamine and histidine [64-67], as well as lipids-related metabolites like fatty acids [64, 66, 68], sphingolipids [67, 69-71], phospholipids [12, 64, 72], and acylcarnitines [12, 73]. These empirical evidences support impaired mitochondrial function [13], glucose metabolism dysregulation [74], and altered lipid metabolism [27, 75] as the defining characteristics of AD patients.

Aging is known to be an important risk factor for AD, and both aging and AD involve perturbed metabolism [76]. Through metabolomic studies, researchers could gain a deeper understanding of the connection between aging and AD [76, 77]. As shown in this review, aging and AD have many shared metabolomic biomarkers, for example, RONS, amino acids, and lipids-related metabolites. In a recent study, Xie et al. [78] specifically investigated the underlying metabolomic mechanism shared between aging and AD. They first identified ten metabolic pathways for aging by analyzing the MS metabolomic data of 183 patients through correlation analysis, random forests and enrichment analysis [78]. Next, the authors performed a systematic review of AD-related metabolomic studies, curating 49 different metabolic pathways [78]. Three pathways, purine metabolism, arginine and proline metabolism, as well as the tricarboxylic acid cycle, were found to be shared between aging and AD [78]. Purine metabolism is closely related to the pentose phosphate pathway and in turn, glycolysis [77]. Previous studies found that excessive glycolysis may accelerate aging onset [79] and dysfunctional glycolysis may be a causative factor of AD [80]. Arginine is an amino acid essential for nitric oxide (NO) synthesis, whose impairment was found to contribute to aging [81], whereas proline is known to help reduce oxidative stress and improve NO availability [81]. In summary, previous studies have identified redox reactions, glycolysis, and lipid metabolism as potential metabolomic links between aging and AD.

Cognitive decline over time is a natural process of aging [82]. However, certain abilities such as vocabulary should be resilient to brain aging [83] but can be impaired in AD patients. Therefore, apart from examining the metabolomic links between aging and AD, it is important to investigate the metabolomic divergence between natural age-related cognitive decline (ARCD) and AD. Motsinger-Reif et al. [84] fitted stepwise logistic regression models to assess the ability of metabolite markers to discriminate between 40 AD patients and 38 cognitively normal controls over 65 years old. Through cross validation, the metabolites 15-65.533 and 8-93.65 were found to be the most consistent predictors [84]. Although their identities are unknown, the authors pinpointed known metabolomic markers significantly correlated with these metabolites such as methionine, glutathione, kynurenine, and indole-3-propionic acid [84]. These results suggest that glutathione synthesis, tryptophan metabolism, and oxidative stress may be involved in the pathogenesis of AD [84]. Hunsberger et al. [85] analyzed the metabolomic data of AD and control (including ARCD) mice through PCA, ANOVA, pathway analysis, and correlation analysis. They identified a difference in the metabolomic profiles of AD and ARCD mice, which was more prominent in the prefrontal cortex and the hippocampus of the right hemisphere [85]. It was found that AD mice exhibit altered protein synthesis (phenylalanine, tyrosine and tryptophan biosynthesis), amino acid catabolism (arginine and proline metabolism; alanine, aspartate, and glutamate metabolism), and histidine metabolism pathways in the prefrontal cortex [85]. On the other hand, the differentially abundant metabolites between the hippocampi of AD and control mice mainly belong to protein synthesis (aminoacyl-tRNA biosynthesis) and oxidative stress (glutathione metabolism; glyoxylate and dicarboxylate metabolism) pathways [85]. In summary, previous research found that the metabolomic divergences between ARCD and AD may be related to protein synthesis, amino acid metabolism, and oxidative stress pathways.

## 4. Multi-omics approaches

Advances in high-throughput technologies for examining genome-wide multi-level omics have enabled researchers to perform integrative analysis of different types of omics data, such as genomic, epigenomic, transcriptomic, proteomic, metabolomic, and microbiomic data [86]. Multi-omics analysis could provide a holistic view of complex traits, elucidating their potential causative factors [15, 86]. There exists an increasing number of multi-omics studies for delineating the molecular taxonomy of aging and AD. It is also worth noting that there exist different online multi-omics resources for studying aging and AD, for example, the Aging Atlas [87], a database hosting aging-related genomic, epigenomic, transcriptomic, proteomic, and metabolomic data, and ADAS-viewer [88], a web server for analyzing genomic, epigenomic, and transcriptomic data from three independent AD cohorts.

To study the molecular changes in aging, Nie et al. [89] used a multi-omics approach to estimate the biological ages (BAs) of human organs and systems utilizing genomic, metabolomic, and microbiomic data from a healthy population cohort of 4,066 Chinese individuals. To construct the BAs, the Klemera and Doudal algorithm [90] was used – after regressing chronological age (CA) on all the available biomarkers, BA was estimated as the linearly best fitted value of CA. The authors found that organs and systems age at diverse rates, and BA can be used for the prediction of mortality and organ-specific disorders such as fatty liver disease [89]. Moreover, the BAs of different human organs and systems were found to be associated with diverse metabolomic pathways [89]. For example, the glutathione metabolism pathway is specifically associated with renal BA, and the fatty acid metabolism pathway is only significantly associated with liver BA [90]. Ahadi et al. [91] performed deep multi-omics longitudinal profiling of 106 healthy individuals through enrichment and regression analyses. The cohort was followed quarterly for up to four years [91]. In each visit, genomic, transcriptomic, metabolomic, and microbiomic data were collected and analyzed [91]. The authors pinpointed 184 molecules and microbes significantly correlated with age, and 160 out of the 184 (86.96%) were found to be metabolites, confirming the strong relationship between aging and metabolism [91]. Among the 160 metabolites, 40% were lipids, 18% were xenobiotics (hippuric acid, 2-aminophenol sulfate and quinic acids), and 16% were amino acids [91], possibly indicating the central role of lipid metabolism in aging. The authors next defined four types of aging patterns (ageotypes) based on the molecular pathways that changed over time in a given individual, which may be useful in tracking and intervening the aging process [91].

AD is known to be heterogenous, and different AD patients may have distinct subphenotypes which progress differently and require precise treatment [92-94]. To characterize the progression and subtypes of AD, Iturria-Medina et al. [29] integrated the epigenomic, transcriptomic, proteomic, and metabolomic data of 1,863 individuals and performed patient subtype stratification using the multi-omics contrastive trajectory inference (mcTI) algorithm. The mcTI framework consisted of six steps: initial feature selection, dimensionality reduction using contrastive principal component analysis [95], dimensionally reduced modalities aggregation using the similarity network fusion algorithm [96], individual molecular disease progression score (mDPS) calculation based on network metrics, patient subtyping using cross-validated expectation-maximization algorithm, and finally, subtype significance evaluation through permutation testing [29]. Three AD subtypes were identified, each having a different set of differentially abundant metabolites [29]. The first subtype, which had a larger number of significant metabolic alterations, presented changes in phosphatidylcholines and amino acids (methionine, histidine, ornithine, and citrulline) [29]. On the other hand, metabolomic changes in subtypes 2 and 3 were mainly related to glutamate, proline, threonine, and valine [29]. These findings indicate that metabolomic changes in AD patients may be AD subtype dependent, and the preventive or treatment interventions of AD may benefit from taking into consideration the subtype of a specific patient [29]. Venugopalan et al. [97] focused on AD stage prediction by building a multi-modal model using human data from the Alzheimer's Disease Neuroimaging Initiative database [98]. They integrated imaging, genetic, and electronic health records (EHRs) data including metabolomic measurements using deep learning and demonstrated that deep and multi-modality frameworks had better predictive performances over shallow or single-modality models [97]. Their results showed that asymmetric dimethylarginine (ADMA) and phosphatidylcholines (PCs) were among the top predictors for the disease stage of AD. ADMA inhibits NO synthase, an enzyme for the synthesis of NO [99], and abnormal NO signalling has been reported in the brains of AD patients [100]. Therefore, ADMA may act as a potential target for AD interventions [101]. PCs are a class of phospholipids, and AD have been frequently reported to be associated with abnormal lipid metabolism [68, 102]. The involvement of PC in AD pathology, as suggested by Venugopalan et al. [97], is supported by previous metabolomic studies [102, 103]. To study the role of brain metabolism in AD pathogenesis, Varma et al. [104] constructed linear mixed models and genome-scale metabolic networks using transcriptomic and metabolomic data. The authors found that despite unchanged free cholesterol levels in AD brains, both de novo cholesterol biosynthesis and catabolism are impacted by AD [104]. Increased non-enzymatic cholesterol catabolism was also observed in AD patients, indicating a shift in metabolomic pathways [104]. Horgusluoglu et al. [105] followed a similar approach to identify AD biomarkers, and constructed metabolite co-expression networks using genomic, transcriptomic, proteomic, and metabolomic data. The study found that acylcarnitines and amines in blood and brain samples are significantly associated with worse AD outcomes and AD endophenotypes, and ABCA1 and CPT1A can be key regulators of acylcarnitines and amino acids in AD [105]. These results provide new evidence of metabolic network failures related to AD pathology and neurodegeneration [105].

## 5. Conclusions and perspectives

Technological advancements in metabolomics have enabled researchers to systematically study the metabolic changes in aging and AD. The aging process was found to be associated with altered amino acid metabolism, lipid metabolism, and redox reactions, whereas AD can be characterized by impaired mitochondrial function as well as dysregulated glucose metabolism and lipid metabolism. Potential metabolomic links between aging and AD include redox reactions, glycolysis, and lipid metabolism. On the other hand, the metabolomic divergences between natural aging and AD may be related to protein synthesis, amino acid metabolism, and oxidative stress pathways. Compared with purely metabolomic approaches, multi-omics studies can better portray the molecular taxonomy of aging and AD. By integrating genomic, epigenomic, transcriptomic, proteomic, and/or microbiomic data with metabolomic data, researchers have successfully constructed metabolic networks related to aging and AD pathology, identified key regulators of metabolomic biomarkers, and identified aging and AD subtypes potentially useful for clinical practice.

Moving forward, many promising directions of research remain to be explored. Spatial transcriptomics provides us with a spatial perspective of the studied tissue in addition to gene expression data [106, 107], and a recent spatial transcriptomics study identified the prominence of a plaque-induced gene network involving oxidative stress in late AD patients [108]. Future multi-omics studies may benefit from integrating metabolomic data with spatial transcriptomics data to better understand the metabolomics of the aging mechanism and AD pathology in the tissue context. Although many metabolomic biomarkers of AD have been identified, whether the metabolites are causal to the pathogenesis of AD often remains unknown. Mendelian randomization can be utilized to assess the causality between the identified metabolomic features and AD [109, 110]. In summary, the rapidly expanding field of metabolomics offers great promise for characterizing the underlying metabolomic continuum of aging and AD, which will be pertinent to precision medicine applications in a rapidly aging world.